\documentclass[aps,pra,showpacs,twocolumn]{revtex4-1}
\usepackage{amssymb}
\usepackage{amsmath}
\usepackage{graphicx}
\usepackage{epsfig}

\begin{document}

\title{Long-range entangled zero-mode state in a non-Hermitian lattice}
\author{S. Lin,${}^1$ X. Z. Zhang,${}^2$ C. Li,${}^1$ and Z. Song,${}^1$}
\email{songtc@nankai.edu.cn}
\affiliation{${\ }^1$School of Physics, Nankai University, Tianjin 300071, China \\
${\ }^2$College of Physics and Materials Science, Tianjin Normal University,
Tianjin 300387, China}

\begin{abstract}
In contrast to a Hermitian system, in which zero modes are usually
degenerate and localized edge state in the thermodynamic limit, the zero
mode of a finite non-Hermitian system can be a single nontrivial long-range
entangled state at the exceptional point (EP). In this work, we demonstrate
this feature with a concrete example based on exact solutions. Numerical
simulations show that the entangled state can be generated through the
dynamic process with a high fidelity.
\end{abstract}

\pacs{11.30.Er, 03.67.Bg, 73.20.At}
\maketitle

%11.30.Er    Charge conjugation, parity, time reversal, and other discrete symmetries
%03.67.Bg	  Entanglement production and manipulation (for entanglement in Bose-Einstein condensates, see 03.75.Gg)
%73.20.At	  Surface states, band structure, electron density of states
%73.20.-r	Electron states at surfaces and interfaces
%03.65.-w Quantum mechanics

\section{Introduction}

Non-Hermitian systems make many things possible including quantum phase
transition occurred in a finite system \cite%
{Znojil1,Znojil2,Bendix,LonghiPRL,LonghiPRB1,Jin1,Znojil3,LonghiPRB2,LonghiPRB3,Jin2,Joglekar1,Znojil4,Znojil5,Zhong,Drissi,Joglekar2,Scott1,Joglekar3,Scott2,Tony}%
, unidirectional propagation and anomalous transport \cite%
{LonghiPRL,Kulishov,LonghiOL,Lin,Regensburger,Eichelkraut,Feng,Peng,Chang},
invisible defects \cite{LonghiPRA2010,Della,ZXZ}, coherent absorption \cite%
{Sun} and self sustained emission \cite%
{Mostafazadeh,LonghiSUS,ZXZSUS,Longhi2015,LXQ}, loss-induced revival of
lasing \cite{PengScience}, as well as laser-mode selection \cite%
{FengScience,Hodaei}. Such kinds of novel phenomena can be traced to the
existence of exceptional or spectral singularity points. Exploring novel
quantum states in non-Hermitian systems becomes an attractive topic. It is
well known that a midgap state must be a localized state without the
long-range correlation. Two degenerate\ edge states can always construct two
long-range correlated states.\ However, it is hard to generate a single
long-range correlated state in practice due to the degeneracy. On the other
hand, a prepared state is fragile due to the interaction with the
environment. A gap protected long-range correlated state is desirable in
many fields. For instance, in quantum information science, it is a crucial
problem to develop techniques for generating entanglement among stationary
qubits, which play a central role in applications \cite{Ekert, Deutsch,
Bennett}.

\begin{figure}[tbp]
\includegraphics[ bb=68 319 485 656, width=0.41\textwidth, clip]{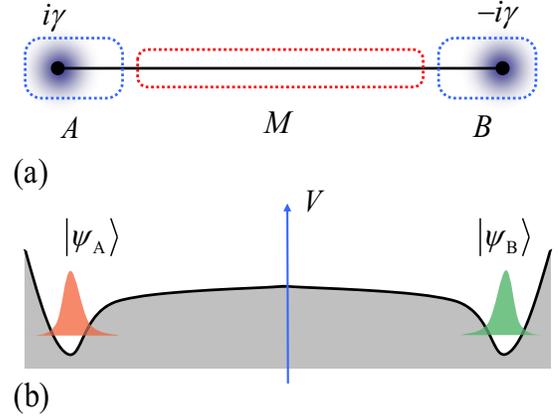}
\caption{(color online). Schematics for a dimerized chain with two end
imaginary potentials as a quantum entangler. (a) The whole chain is
decomposed into three parts. A and B are both $N_0$-site, $M$ is $N-2N_0$
site. At exceptional point (EP), there is a single zero-mode state, which is an approximately
maximal entangled state for the spatial modes of parties A and B. (b)
Schematic illustration of the zero-mode state. The chain system acts as a
double well potential, which has two spatial modes $\left\vert \protect\psi %
_{\mathrm{A}}\right\rangle $ and $\left\vert \protect\psi _{\mathrm{B}%
}\right\rangle $. The zero-mode state is similar to the ground state of the
potential well, being an entangled state of the two spatial modes.}
\label{fig1}
\end{figure}

In this paper, we consider whether it is possible to find a midgap state
with nontrivial long-range correlation in a non-Hermitian system. We revisit
the system of a tight-binding chain with two conjugated imaginary potentials
at the end sites, which has received many studies \cite%
{Bendix,Jin1,Joglekar1,Joglekar2,Joglekar3,Longhi2013,Ganainy,Makris,Guo,Ruter,Regensburger,ChenS}%
. Based on the exact results, we show that there exists a single zero-mode
state with long-range correlation. This system acts as a double well
potential, which has two spatial modes (see Fig. \ref{fig1}). We propose a
scheme to generate a midgap entangled state through the dynamic process in a
finite non-Hermitian system. Numerical simulations show that the entangled
zero-mode state can be achieved with a high fidelity via the time evolution
of an easily prepared initial state.

The remainder of this paper is organized as follows. In Sec. \ref{Model and
coalescing zero-mode}, we present a non-Hermitian chain and zero-mode
solutions. Sec. \ref{Hermitian correspondence for zero-mode state} reveals
the implication of the zero-mode solution. Sec. \ref{Mode entanglement}
demonstrates the mode entanglement in the zero-mode state. Sec. \ref%
{Generation of entanglement} devotes to the scheme for the generation of
long-range entanglement. Finally, we present a summary and discussion in
Sec. \ref{Summary}.

\section{Model and coalescing zero-mode}

\label{Model and coalescing zero-mode}

We consider a non-Hermitian model with complex boundary potentials and its
Hamiltonian is%
\begin{eqnarray}
H &=&H_{0}+H_{\gamma },  \notag \\
H_{0} &=&-\kappa \sum_{j=1}^{N/2}a_{2j-1}^{\dagger
}a_{2j}-\sum_{j=1}^{N/2-1}a_{2j}^{\dagger }a_{2j+1}+\mathrm{H.c.,} \\
H_{\gamma } &=&i\gamma a_{1}^{\dagger }a_{1}-i\gamma a_{N}^{\dagger }a_{N},
\notag
\end{eqnarray}%
where $a_{l}^{\dag }$ is the creation operator of the boson (or fermion) at $%
l$th site. There are two sets of hopping integrals with strength\textbf{\ }$%
\kappa $ and $1$, respectively. For simplicity, we only consider the case
with $N/2=$ even. The Hamiltonian $H$ has $\mathcal{PT}$ symmetry, i.e., $[H,%
\mathcal{PT}]=0$, where $\mathcal{P}$ and $\mathcal{T}$\ represent the
space-reflection operator, or parity operator and the time-reversal
operator, respectively. The effects of $\mathcal{P}$ and $\mathcal{T}$\ on a
discrete system are
\begin{equation}
\mathcal{T}i\mathcal{T}=-i\text{, }\mathcal{P}a_{l}^{\dag }\mathcal{P}%
=a_{N+1-l}^{\dag }.
\end{equation}%
Since the discovery of Bender and colleagues in 1998 that a non-Hermitian
Hamiltonian having simultaneous parity-time $\mathcal{PT}$\ symmetry has an
entirely real quantum-mechanical energy spectrum \cite{Bender}, there has been an intense
effort to establish a $\mathcal{PT}$-symmetric quantum theory as a complex
extension of the conventional quantum mechanics. The reality of the spectra
is responsible to the $\mathcal{PT}$\ symmetry. If all the eigenstates of
the Hamiltonian are also eigenstates of $\mathcal{PT}$, then all the
eigenvalues are strictly real and the symmetry is said to be unbroken.
Otherwise, the symmetry is said to be spontaneously broken. On the other
hand, the unitary time-evolution of wave functions in a non-Hermitian
Hamiltonian could also be guaranteed when redefining the biorthogonal inner
product instead of the Dirac inner product. However, when the system is\
under the EP, the norm of biorthogonal inner product for coalescing
eigenstates is zero.

In the case of $\kappa =1$, it has been shown in Ref. \cite{Jin1} that this
model exhibits two phases, an unbroken symmetry phase with a purely real
energy spectrum when the potentials are in the region $\gamma <1$ and a
spontaneously-broken symmetry phase with $N-2$ real and $2$ imaginary
eigenvalues when the potentials are in the region $\gamma >1$. The EP occurs
at $\gamma =1$, at which two zero-energy eigenstates coalesce. In the case
of $\kappa \neq 1$, the model is systematically investigated in Ref. \cite%
{ChenS}.

In this work, we revisit this model from another perspective, focusing on
the property of coalescing zero-mode state based on exact solutions. We
start with the case of $\gamma =0$\ and $0<\kappa <1$. According
to bulk-boundary correspondence, there are two zero modes at the middle of
the band gap for infinite $N$, characterizing the topology of the band \cite%
{Ryu,Ganeshan}. These two states are all called edge modes, since they are
localized states, with the particle probability distributing at the ends of
the chain. For finite $N$, two zero modes split near the zero energy. We are
interested in what happens when the imaginary potentials $\pm i\gamma $\ are
switched on. It is a little bit complicated to get and analyze the exact
solutions. Numerical simulation shows that the EP occurs at a certain value
of $\gamma $, at which the two midgap states coalesce to a single
eigenstate. Fortunately, we can demonstrate this point by a simple exact
result.\

\begin{figure}[tbp]
\includegraphics[ bb=87 282 504 574, width=0.47\textwidth, clip]{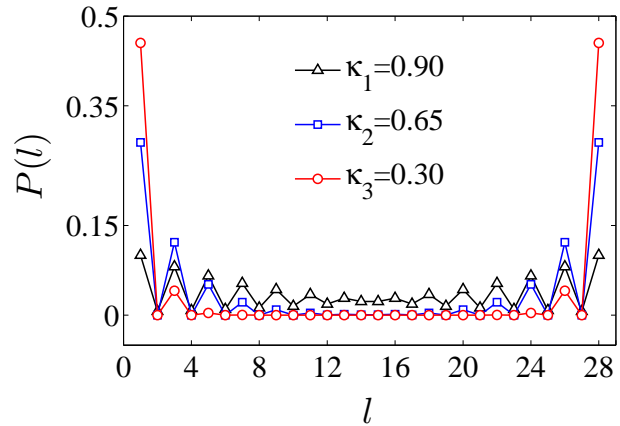}
\caption{(color online). Profiles of the Dirac probability distribution of
the zero-mode states for systems at their corresponding EPs. We consider an $%
N=28$ chain with typical values of $\protect\kappa $ with critical $\protect%
\gamma _{\mathrm{c}}=\protect\kappa ^{N/2}$, respectively.}
\label{fig2}
\end{figure}

A straightforward derivation shows that when taking
\begin{equation}
\gamma =\gamma _{\mathrm{c}}=\kappa ^{N/2},  \label{Gamma_Condition}
\end{equation}%
the state defined as $\left\vert \psi _{\mathrm{zm}}\right\rangle =\psi _{%
\mathrm{zm}}^{\dag }\left\vert \text{\textrm{Vac}}\right\rangle $ is a
zero-mode eigenstate, where state $\left\vert \text{\textrm{Vac}}%
\right\rangle $\ is defined as%
\begin{equation}
\left\vert \text{\textrm{Vac}}\right\rangle =\prod_{j=1}^{N}\left\vert
0\right\rangle _{j},
\end{equation}%
and $\left\vert 0\right\rangle _{j}$\ is the vacuum state of particle
operator $a_{j}^{\dagger }$, i.e.,\textbf{\ }$a_{j}\left\vert 0\right\rangle
_{j}=0$. And field operator is%
\begin{equation}
\psi _{\mathrm{zm}}^{\dag }=\Omega \sum_{j=1}^{N/2}[\left( -\kappa \right)
^{j-1}a_{2j-1}^{\dagger }-i\left( -\kappa \right) ^{N/2-j}a_{2j}^{\dagger }],
\end{equation}%
i.e.,%
\begin{equation}
H\left( \gamma _{\mathrm{c}}\right) \left\vert \psi _{\mathrm{zm}%
}\right\rangle =0.
\end{equation}%
Furthermore, the many-particle state $(\psi _{\mathrm{zm}}^{\dag
})^{n}\left\vert \text{\textrm{Vac}}\right\rangle $ is still a zero-mode
eigenstate. Here $\Omega =\sqrt{\left( 1-\kappa ^{2}\right) /\left(
2-2\kappa ^{N}\right) }$ is the Dirac normalizing constant. Similarly, the
zero-mode state for $H_{\gamma _{\mathrm{c}}}^{\dagger }$ can be constructed
as%
\begin{equation}
\left\vert \eta _{\mathrm{zm}}\right\rangle =\Omega \sum_{j=1}^{N/2}[\left(
-\kappa \right) ^{j-1}a_{2j-1}^{\dagger }+i\left( -\kappa \right)
^{N/2-j}a_{2j}^{\dagger }]\left\vert \text{\textrm{Vac}}\right\rangle ,
\end{equation}%
satisfying
\begin{equation}
\lbrack H\left( \gamma _{\mathrm{c}}\right) ]^{\dagger }\left\vert \eta _{%
\mathrm{zm}}\right\rangle =0.
\end{equation}%
On the other hand, it is easy to check
\begin{equation}
\left\langle \eta _{\mathrm{zm}}\right. \left\vert \psi _{\mathrm{zm}%
}\right\rangle =0,
\end{equation}%
and
\begin{figure}[tbp]
\includegraphics[ bb=90 433 433 694, width=0.40\textwidth, clip]{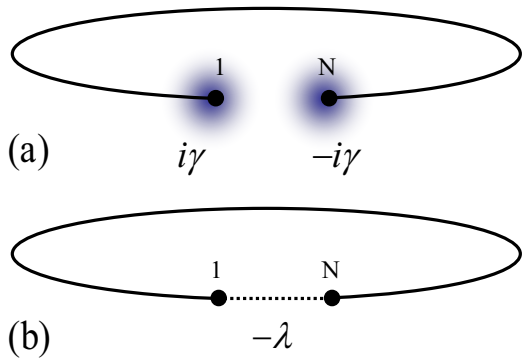}
\caption{(color online). Schematics for the Hermitian correspondence of the
zero-mode state. (a) $\mathcal{PT}$ symmetric non-Hermitian chain, (b) $%
\mathcal{P}$ symmetric Hermitian ring. When the on-site imaginary potentials
$\pm i\protect\gamma $ and the boundary hopping $\protect\lambda $ satisfy
the zero-mode conditions in Eq. (\protect\ref{Gamma_Condition}) and (\protect
\ref{Lambda_Condition}), the two systems share a same zero-mode state. }
\label{fig3}
\end{figure}
\begin{equation}
\left\vert \eta _{\mathrm{zm}}\right\rangle =i\mathcal{P}\left\vert \psi _{%
\mathrm{zm}}\right\rangle ,
\end{equation}%
which indicate that the coalescing levels typically produce not only
eigenstates but also adjoint states. In the context of non-Hermitian quantum
mechanics, the adjoint states refer to the eigenstates of $H^{\dag }$, which
always share the same spectrum with those of $H$. Both the eigenstates of $H$%
\ and $H^{\dag }$\ could construct the biorthogonal inner product instead of
the Dirac inner product in order to guarantee the unitary time-evolution of
wave functions in a non-Hermitian Hamiltonian. When the system is under the
EP, two eigenstates of $H$\ coalesce and the norm of biorthogonal inner
product for coalescing eigenstates is zero. Meanwhile, it is a natural
result that two eigenstates of $H^{\dag }$\ also coalesce due to the fact $%
H^{\dag }=H(\gamma \rightarrow -\gamma )$. Here the two coalescing
eigenstates of $H$($H^{\dag }$) would become only one eigenstate in the
non-Hermitian system while there are two-fold degenerate states in the
Hermitian system that we will consider in the next section. The existence of two-fold degenerate entangled states in
the Hermitian system would degrade the efficiency of the entangled state in
practice. While, there is a single midgap entangled state in the
non-Hermitian system, which is robust.

Based on these facts we conclude that the zero-mode state $\left\vert \psi _{%
\mathrm{zm}}\right\rangle $\ is a coalescing state and the EP occurs at $%
\gamma _{\mathrm{c}}$. The exact wave function of $\left\vert \psi _{\mathrm{%
zm}}\right\rangle $ clearly indicates that it is a\ superposition of two
parts with nonzero amplitudes only located at even or odd sites,
respectively. It will be shown that such two parts, which consist of even
and odd sites respectively, have a close relation to the standard edge
states of a Hermitian chain in the thermodynamic limit.

In our work, we explore the quantum correlation, entanglement of a midgap
state for a non-Hermitian Hamiltonian. Unlike the dynamics, such the feature
of a state has no memory about its own history. One cannot detect how the
state was prepared, via a Hermitian or non-Hermitian system. In contrast,
the metric of biorthogonal inner product in non-Hermitian quantum mechanics
is dependent of Hamiltonian. Meanwhile, when the system is under the EP, the
norm of biorthogonal inner product for coalescing eigenstates is zero, which
is the characteristic of a non-Hermitian system. On the other hand, the
entanglement is measured via local Dirac probability in experiments. In Fig. %
\ref{fig2}, we plot the profiles of the Dirac probability distribution of
the zero-mode states in a $28$-site chain at EPs where all the zero-mode
states have been normalized in Dirac inner product. Here the local Dirac
probability is defined as $P\left( l\right) =\left\vert \left\langle
l\right. \left\vert \psi _{\mathrm{zm}}\right\rangle \right\vert ^{2}$\ with
$\left\vert l\right\rangle =a_{l}^{\dagger }\left\vert \mathrm{Vac}%
\right\rangle $. We see that a smaller $\kappa $\ requires a smaller $\gamma
_{\mathrm{c}}$, leading to more localized probability distribution\ and
bigger gap. It is the key feature of the non-Hermitian zero-mode state,
which should have wild applications in many aspects. We will demonstrate
this point explicitly in the following sections.

\section{Hermitian correspondence for zero-mode state}

\label{Hermitian correspondence for zero-mode state}

In this section, we investigate the property of the zero-mode state. We
consider a Hermitian model on the same lattice with $H$ and its Hamiltonian
is%
\begin{eqnarray}
H_{\mathrm{hc}} &=&H_{0}+H_{\lambda }, \\
H_{\lambda } &=&-\lambda a_{1}^{\dagger }a_{N}+\mathrm{H.c.}.
\end{eqnarray}%
The Hamiltonian $H_{\mathrm{hc}}$ has $\mathcal{P}$ symmetry, i.e., $[H_{%
\mathrm{hc}},\mathcal{P}]=0$. Then all the eigenstates should have
reflection symmetry. We still investigate the midgap state through a simple
exact solution. In Fig. \ref{fig3}, the two systems $H$\ and $H_{\mathrm{hc}%
} $ are schematically illustrated.

A straightforward derivation shows that when taking
\begin{equation}
\lambda =\lambda _{\mathrm{c}}=\kappa ^{N/2},  \label{Lambda_Condition}
\end{equation}%
the state is defined as $\left\vert \psi _{\sigma }\right\rangle =\psi
_{\sigma }^{\dag }\left\vert \text{\textrm{Vac}}\right\rangle $ with $\sigma
=\pm $, and
\begin{equation}
\psi _{\pm }^{\dag }=\Omega \sum_{j=1}^{N/2}[\left( -\kappa \right)
^{j-1}a_{2j-1}^{\dagger }\pm i\left( -\kappa \right) ^{N/2-j}a_{2j}^{\dagger
}],  \label{ZM_Hermitian}
\end{equation}%
is the field operator of zero-mode eigenstate, i.e.,
\begin{equation}
H\left( \lambda _{\mathrm{c}}\right) \left\vert \psi _{\pm }\right\rangle =0.
\end{equation}%
It is easy to find that%
\begin{equation}
\left\vert \psi _{\mathrm{zm}}\right\rangle =\left\vert \psi
_{-}\right\rangle ,\left\vert \eta _{\mathrm{zm}}\right\rangle =\left\vert
\psi _{+}\right\rangle .
\end{equation}%
The relation between zero-mode states of the non-Hermitian Hamiltonian $H$
and the Hermitian one $H_{\mathrm{hc}}$ is very clear, that is, the two
Hamiltonians share a same zero-mode state.

Furthermore, one can obtain the exact wave function for the standard zero
modes\textbf{\ }by simply taking $N\rightarrow \infty $, i.e., $\lambda _{%
\mathrm{c}}=\lim_{N\rightarrow \infty }\kappa ^{N/2}=0$. Then our Hermitian
model would become a standard SSH chain and accordingly the standard edge
states can be defined as $\left\vert \psi _{Y}\right\rangle =\psi _{Y}^{\dag
}\left\vert \text{\textrm{Vac}}\right\rangle $\ with $Y=L$, $R$, where%
\textbf{\ }%
\begin{eqnarray}
\psi _{\mathrm{L}}^{\dag } &=&\sqrt{2}\Omega \sum_{j=1}^{\infty }\left(
-\kappa \right) ^{j-1}a_{2j-1}^{\dagger }, \\
\psi _{\mathrm{R}}^{\dag } &=&\sqrt{2}\Omega \sum_{j=1}^{\infty }\left(
-\kappa \right) ^{N/2-j}a_{2j}^{\dagger },
\end{eqnarray}%
with $\left\vert \psi _{\mathrm{L}}\right\rangle =\lim_{N\rightarrow \infty
}\left( \left\vert \psi _{+}\right\rangle +\left\vert \psi _{-}\right\rangle
\right) /\sqrt{2}$\ and $\left\vert \psi _{\mathrm{R}}\right\rangle
=-i\lim_{N\rightarrow \infty }\left( \left\vert \psi _{+}\right\rangle
-\left\vert \psi _{-}\right\rangle \right) /\sqrt{2}$.

Interestingly, one can see that the profile of the wave functions are
unchanged. Then we get the conclusion that wave functions $\left\vert \psi _{%
\mathrm{zm}}\right\rangle $, $\left\vert \eta _{\mathrm{zm}}\right\rangle ,$
and $\left\vert \psi _{\pm }\right\rangle $, can be obtained directly by the
truncations of the standard zero-mode wave functions.

It appears an interesting \textquotedblleft coincidence\textquotedblright\ that the zero-mode states in Eq. (\ref%
{ZM_Hermitian}) are identical to those of the non-Hermitian Hamiltonian.
Here we would like to give a further explanation on the essential
of the \textquotedblleft coincidence\textquotedblright. We note that in
both Hermitian and non-Hermitian lattices, the sub-Hamiltonians in the
region far from the boundary are identical, i.e.,%
\begin{equation}
H_{\mathrm{mid}}=-\kappa \sum_{j}a_{2j-1}^{\dagger
}a_{2j}-\sum_{j}a_{2j}^{\dagger }a_{2j+1}+\mathrm{H.c.}.
\end{equation}%
And in this region the single-particle wave function can be expressed as%
\begin{eqnarray}
\left\vert \psi _{k}\right\rangle
&=&\sum_{j}(A_{k}e^{ikj}+B_{k}e^{-ikj})a_{2j-1}^{\dagger }\left\vert \text{%
\textrm{Vac}}\right\rangle   \notag \\
&&+\sum_{j}(C_{k}e^{ikj}+D_{k}e^{-ikj})a_{2j}^{\dagger }\left\vert \text{%
\textrm{Vac}}\right\rangle ,
\end{eqnarray}%
where $A_{k}$, $B_{k}$, $C_{k}$, and $D_{k}$ are undetermined parameters.
Then the corresponding Schrodinger equation for a zero-mode state has the
forms%
\begin{equation}
\left\{
\begin{array}{c}
-\kappa (A_{k}e^{ikj}+B_{k}e^{-ikj})-\left[ A_{k}e^{ik\left( j+1\right)
}+B_{k}e^{-ik\left( j+1\right) }\right] =0, \\
-\kappa (C_{k}e^{ikj}+D_{k}e^{-ikj})-\left[ C_{k}e^{ik\left( j-1\right)
}+D_{k}e^{-ik\left( j-1\right) }\right] =0,%
\end{array}%
\right.
\end{equation}%
which lead to%
\begin{equation}
\left\{
\begin{array}{c}
\left( \kappa +e^{ik}\right) A_{k}e^{ikj}+\left( \kappa +e^{-ik}\right)
B_{k}e^{-ikj}=0, \\
\left( \kappa +e^{-ik}\right) C_{k}e^{ikj}+\left( \kappa +e^{ik}\right)
D_{k}e^{-ikj}=0,%
\end{array}%
\right.
\end{equation}%
or the compact form
\begin{equation}
\left(
\begin{array}{cccc}
h_{11} & 0 & 0 & 0 \\
0 & h_{22} & 0 & 0 \\
0 & 0 & h_{33} & 0 \\
0 & 0 & 0 & h_{44}%
\end{array}%
\right) \left(
\begin{array}{c}
A_{k} \\
B_{k} \\
C_{k} \\
D_{k}%
\end{array}%
\right) =0,
\end{equation}%
with%
\begin{eqnarray}
h_{11} &=&h_{44}=\kappa +e^{ik}, \\
h_{22} &=&h_{33}=\kappa +e^{-ik}.
\end{eqnarray}%
The existence of the solution requires that%
\begin{equation}
\det \left(
\begin{array}{cccc}
h_{11} & 0 & 0 & 0 \\
0 & h_{22} & 0 & 0 \\
0 & 0 & h_{33} & 0 \\
0 & 0 & 0 & h_{44}%
\end{array}%
\right) =0,
\end{equation}%
or%
\begin{equation}
\left( \kappa +e^{ik}\right) \left( \kappa +e^{-ik}\right) =0.
\end{equation}%
In our work, we only consider the case with $0<\kappa <1$. Then the possible solution of $k$\ is
\begin{equation}
k=\pi -i\ln \kappa \text{ or }k=\pi +i\ln \kappa ,
\end{equation}%
which describes an evanescent wave in accord with the edge state at the
midgap.

This fact indicates that both Hermitian and non-Hermitian systems share a
common $k$\ and therefore the main form of their wave functions is similar.
The parameters $A_{k}$, $B_{k}$, $C_{k}$, and $D_{k}$\ or the relations
between them could be finally determined by the boundary and normalization
conditions. Based on the above analysis, we could achieve two degenerate
solutions for the Hermitian system, while only one solution for the
non-Hermitian system (for the zero-mode state of $H^{\dagger }$, we can only
follow the same method above after taking $\gamma \rightarrow -\gamma $ due
to the reason $H^{\dag }=H(\gamma \rightarrow -\gamma )$). In this sense,
the \textquotedblleft coincidence\textquotedblright\ becomes natural if
suitable boundary conditions for the two systems are chosen.

\section{Mode entanglement}

\label{Mode entanglement}

Entanglement is an intriguing characteristic of quantum mechanics and is
fundamentally different from any correlation known in classical physics. It
would be interesting for both quantum information and condensed matter if
one could generate particular entangled states in a controlled manner. The
typical entanglement refers to a two-particle system. However, it has been
demonstrated that the mode entanglement of a single particle can be used for
dense coding and quantum teleportation despite the superselection rule \cite%
{Heaney}. In the following, we study the mode entanglement in the zero-mode
state.

First of all, we define two spacial-mode states $\left\vert \psi _{\mathrm{A}%
}\right\rangle =\psi _{\mathrm{A}}^{\dag }\left\vert \text{\textrm{Vac}}%
\right\rangle $\ and $\left\vert \psi _{\mathrm{B}}\right\rangle =\psi _{%
\mathrm{B}}^{\dag }\left\vert \text{\textrm{Vac}}\right\rangle $, i.e.,
\begin{eqnarray}
\left\vert \psi _{\mathrm{A}}\right\rangle  &=&\Omega
_{0}\sum_{j=1}^{N_{0}/2}\left( -\kappa \right) ^{j-1}a_{2j-1}^{\dagger
}\left\vert \text{\textrm{Vac}}\right\rangle ,  \label{Spacial_A} \\
\left\vert \psi _{\mathrm{B}}\right\rangle  &=&-i\Omega _{0}\sum_{j=N_{%
\mathrm{B}}}^{N/2}\left( -\kappa \right) ^{N/2-j}a_{2j}^{\dagger }\left\vert
\text{\textrm{Vac}}\right\rangle ,  \label{Spacial_B}
\end{eqnarray}%
where $\Omega _{0}=\sqrt{\left( 1-\kappa ^{2}\right) /\left( 1-\kappa
^{N_{0}}\right) }$\ is the Dirac normalizing constant and $N_{\mathrm{B}%
}=\left( N-N_{0}+2\right) /2$. $\psi _{X}^{\dag }$\ is the field operator
for spatial mode $X=A$, $B$,\ i.e., $\psi _{\mathrm{A}}^{\dag }\left\vert
\text{\textrm{Vac}}\right\rangle =\left\vert 1\right\rangle _{\mathrm{A}%
}\left\vert 0\right\rangle _{\mathrm{B}}$\ and $\psi _{\mathrm{B}}^{\dag
}\left\vert \text{\textrm{Vac}}\right\rangle =\left\vert 0\right\rangle _{%
\mathrm{A}}\left\vert 1\right\rangle _{\mathrm{B}}$. Here%
\begin{eqnarray}
\left\vert 1\right\rangle _{\mathrm{A}} &=&\psi _{\mathrm{A}}^{\dag
}\left\vert 0\right\rangle _{\mathrm{A}},\left\vert 1\right\rangle _{\mathrm{%
B}}=\psi _{\mathrm{B}}^{\dag }\left\vert 0\right\rangle _{\mathrm{B}}, \\
\left\vert 0\right\rangle _{\mathrm{A}} &=&\prod_{j=1}^{N/2}\left\vert
0\right\rangle _{2j-1},\left\vert 0\right\rangle _{\mathrm{B}%
}=\prod_{j=1}^{N/2}\left\vert 0\right\rangle _{2j}.
\end{eqnarray}%
And the spacial mode states defined in Eqs. (\ref{Spacial_A}) and (\ref%
{Spacial_B}) are introduced to characterize the entanglement feature of the
zero-mode states, the formalism of which is adopted and studied in Ref. \cite%
{Heaney}. And these two spacial mode states could construct a target
entangled state $(\left\vert \psi _{\mathrm{A}}\right\rangle +\left\vert
\psi _{\mathrm{B}}\right\rangle )/\sqrt{2}=(\left\vert 1\right\rangle _{%
\mathrm{A}}\left\vert 0\right\rangle _{\mathrm{B}}+\left\vert 0\right\rangle
_{\mathrm{A}}\left\vert 1\right\rangle _{\mathrm{B}})/\sqrt{2}$, which can
be used to judge the degree of entanglement of the zero-mode state of $H$.

Taking\textbf{\ }$N_{0}=N$, the zero-mode state can be written as%
\begin{equation}
\left\vert \psi _{\mathrm{zm}}\right\rangle =\frac{1}{\sqrt{2}}(\left\vert
1\right\rangle _{\mathrm{A}}\left\vert 0\right\rangle _{\mathrm{B}%
}+\left\vert 0\right\rangle _{\mathrm{A}}\left\vert 1\right\rangle _{\mathrm{%
B}})
\end{equation}%
which is a maximal entanglement. When we measure the state of the
subsystem $A$ $(B)$, it will collapse to either $\left\vert 1\right\rangle _{%
\mathrm{A}}\left\vert 0\right\rangle _{\mathrm{B}}$\ or $\left\vert
0\right\rangle _{\mathrm{A}}\left\vert 1\right\rangle _{\mathrm{B}}$, i.e.,
when $\left\vert \psi _{\mathrm{A}}\right\rangle $\ is detected at $A$,
party $B$ must collapse to empty state $\left\vert 0\right\rangle _{\mathrm{B%
}}$. However, such an entangled state is not useful for teleportation in
quantum information processing, since the particle probability may not
distribute locally.

An interesting fact is that, for $N_{0}\ll N$, but small $\kappa $, we still
have
\begin{equation}
\left\vert \psi _{\mathrm{zm}}\right\rangle \approx \frac{1}{\sqrt{2}}%
(\left\vert 1\right\rangle _{\mathrm{A}}\left\vert 0\right\rangle _{\mathrm{B%
}}+\left\vert 0\right\rangle _{\mathrm{A}}\left\vert 1\right\rangle _{%
\mathrm{B}}),
\end{equation}%
i.e., the zero-mode state is a highly entangled state for the localized
spacial-mode states.

And here we introduce the quantity%
\begin{equation}
\mathcal{F}(N,N_{0},\kappa )=\left\vert \left\langle \psi _{\mathrm{zm}%
}\right\vert \frac{\left\vert 1\right\rangle _{\mathrm{A}}\left\vert
0\right\rangle _{\mathrm{B}}+\left\vert 0\right\rangle _{\mathrm{A}%
}\left\vert 1\right\rangle _{\mathrm{B}}}{\sqrt{2}}\right\vert ,
\end{equation}%
to characterize the entanglement of a zero-mode state for the situation with
parameters $N$, $N_{0}$, and $\kappa $. A simple derivation yields%
\begin{equation}
\mathcal{F}(N,N_{0},\kappa )=\sqrt{\frac{1-\kappa ^{N_{0}}}{1-\kappa ^{N}}},
\end{equation}%
which is close to $\sqrt{1-\kappa ^{N_{0}}}$\ for $N\gg N_{0}$. It\
indicates that a system with $\kappa ^{N_{0}}\ll 1$\ can achieve a perfect
entangled state.

\section{Generation of entanglement}
\label{Generation of entanglement}
\begin{figure}[tbp]
\includegraphics[ bb=44 180 527 601, width=0.48\textwidth, clip]{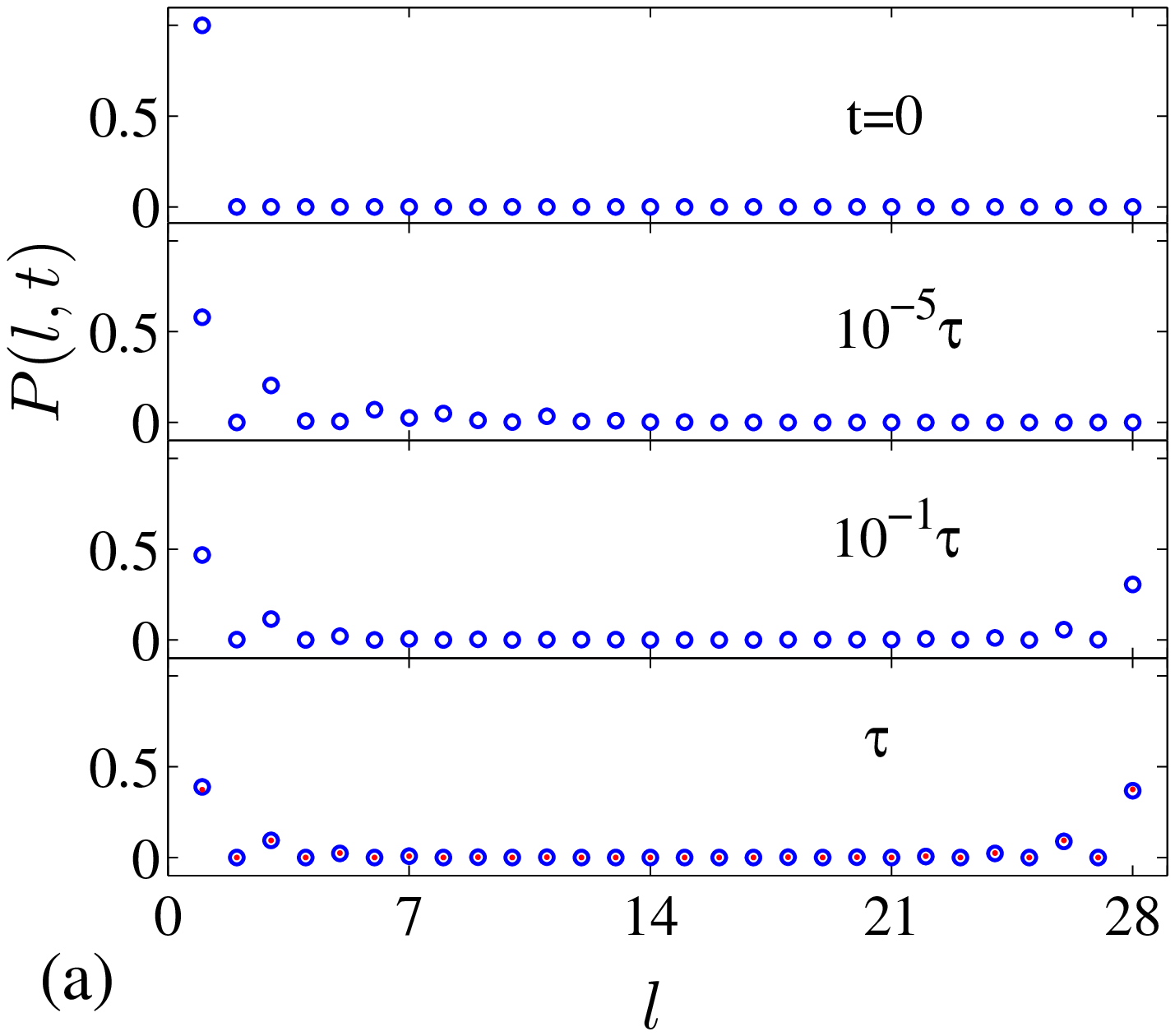} %
\includegraphics[ bb=47 224 530 570, width=0.48\textwidth, clip]{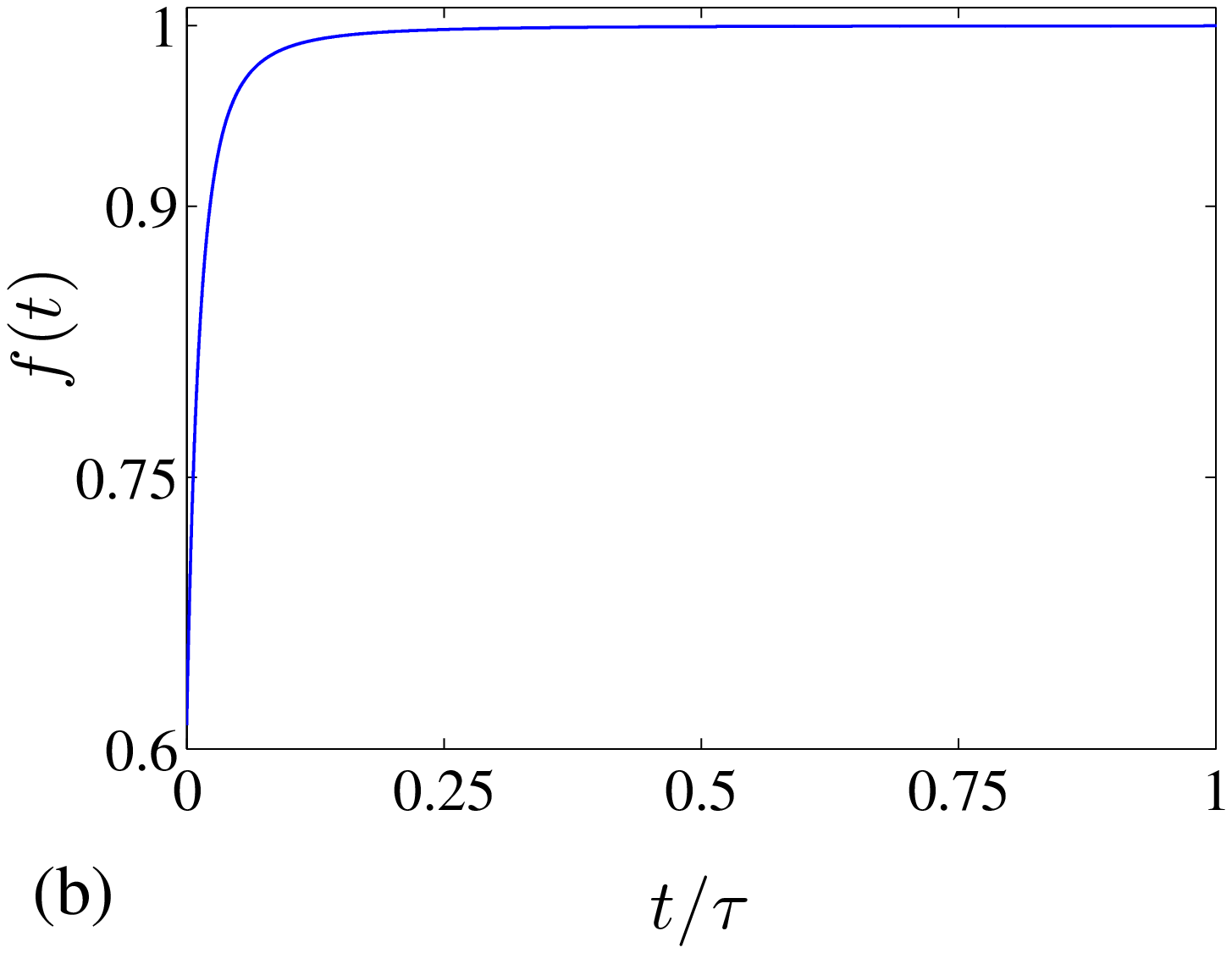}
\caption{(color online). Dynamical process for the generation of the
entangled zero-mode state. (a) The probability distributions $P(l,t)$ for
several instants are obtained by the time evolution of the simplest initial
state $\left\vert 1\right\rangle $ under the Hamiltonian $H$ with $N=28$, $%
\protect\kappa =0.5$, and $\protect\gamma =\protect\gamma _{\mathrm{c}%
}+10^{-10}$. Here the red dot indicates the probability distribution of the
zero-mode state. (b) Plot of the fidelity as a function of time with $%
\protect\tau =8\times 10^{5}.$}
\label{fig4}
\end{figure}
Entanglement naturally exists in the zero-mode state, which is protected by
the band gap. The question in practice is how to create such a state.
Recently, generation scheme of several typical entangled states has been
proposed by the dynamical process\ near the EP of non-Hermitian systems \cite%
{Tony1,Tony2,LC}. The key to the scheme is based on the fact that a
pseudo-Hermitian system has real eigenvalues or conjugate pair complex
eigenvalues \cite{Bender,Ann,JMP1,JPA1,PRL1,JMP2,JPA3,JPA5}. Considering the
situation in this work that the single zero-mode state splits to a pair of
eigenstates with conjugate complex eigenvalues and the $\mathcal{PT}$
symmetry of those eigenstates have been broken, when we tune the potential
to $\gamma =\gamma _{\mathrm{c}}+0^{+}$. This pair of states are very close
to the zero-mode state, in the sense of that they are still long-range
entangled states. A seed state is an initial state consisting of various
eigenstates with their eigenvalues including zero, positive, and negative
imaginary parts, respectively. As time evolution, the amplitude of the state
with positive imaginary part among the eigenvalues will increase
exponentially and suppress that of other components. The target is the final
steady state, which can be regarded as the zero-mode state in the point of
view of Dirac product.

The initial state is taken as $\left\vert \psi \left( 0\right) \right\rangle
=\left\vert 1\right\rangle $, and the evolved state $\left\vert \psi \left(
t\right) \right\rangle $\ is expected to close to the target state $%
\left\vert \psi _{\mathrm{zm}}\right\rangle $ for a sufficient long time.
Whether the Hamiltonian operator $H$\ is Hermitian or not, an evolved vector
$\left\vert \psi \left( t\right) \right\rangle $\ should always be the
solution of the Schrodinger equation%
\begin{equation}
i\frac{\partial }{\partial t}\left\vert \psi \left( t\right) \right\rangle
=H\left\vert \psi \left( t\right) \right\rangle .
\end{equation}%
And the solution has the form%
\begin{equation}
\left\vert \psi \left( t\right) \right\rangle =e^{-iHt}\left\vert \psi
\left( 0\right) \right\rangle ,
\end{equation}%
which can be computed by using a uniform mesh in the time discretization for
method. Of course, by taking the biorthogonal inner product instead of the
Dirac inner product in order to guarantee the unitary time-evolution of the
wave functions in a non-Hermitian Hamiltonian, one can also compute the time
evolution by expanding the initial state in terms of the complete set of
biorthonormal energy eigenstates. We employ the fidelity%
\begin{equation}
f\left( t\right) =\left\vert \left\langle \psi _{\mathrm{zm}}\right\vert
\widetilde{\psi }\left( t\right) \rangle \right\vert ,
\end{equation}%
to characterize the efficiency of the scheme. Here $\left\vert \widetilde{%
\psi }\left( t\right) \right\rangle $\ is the Dirac normalized state of $%
\left\vert \psi \left( t\right) \right\rangle $ to reduce the increasing
norm of $\left\vert \psi \left( t\right) \right\rangle $. In order to
quantitatively evaluate the fidelity and demonstrate the proposed scheme, we
simulate the dynamic process of the state $\left\vert \psi _{\mathrm{zm}%
}\right\rangle $ preparation. We plot the profiles of $\left\vert \widetilde{%
\psi }\left( t\right) \right\rangle $\ at several typical instants and the
fidelity $f\left( t\right) $ in Fig. \ref{fig4}. We see that the
perfect edge state $\left\vert 1\right\rangle $ spreads out rapidly at the
beginning and converges to the target state with a high fidelity after a
long time.

\section{Summary}
\label{Summary}In conclusion, we have studied the midgap zero-mode state in
a non-Hermitian discrete system. The exact result for a dimerized
tight-binding chain with two end imaginary potentials substantiates the
existence of coalescing zero-mode state. Based on the exact results, it is
found that, the zero-mode state exhibits robust long-range spatial mode
entanglement, which is not achievable in a Hermitian system. We have also
investigated the entanglement generation scheme. Numerical simulations
indicate that such a midgap state can be generated by the time evolution of
an easily prepared initial state. Our finding extends the understanding of
non-Hermitian quantum mechanics and should have wide applications in quantum
engineering.\newline
\newline

\acknowledgments We acknowledge the support of the National Basic Research
Program (973 Program) of China under Grant No. 2012CB921900 and CNSF (Grant
No. 11374163).

\end{document}